# Stationary CT Imaging of Intracranial Hemorrhage with Diffusion Posterior Sampling Reconstruction

A Lopez-Montes, T. McSkimming, A. Skeats, C. Delnooz, B. Gonzales, W. Zbijewski, and A. Sisniega

***Abstract*— Diffusion Posterior Sampling (DPS) can be used in Computed Tomography (CT) reconstruction by leveraging diffusion-based generative models for unconditional image synthesis while matching the observations (data) of a CT scan. Of particular interest is its application in scenarios involving sparse or limited angular sampling, where conventional reconstruction algorithms are often insufficient. We developed a DPS algorithm for 3D reconstruction from a stationary CT (sCT) portable brain stroke imaging unit based on a multi-x-ray source array (MXA) of 31 x-ray tubes and a curved area detector. In this configuration, conventional reconstruction – e.g., Penalized Weighted Least Squares (PWLS) with a Huber edge-preserving penalty – suffers from severe directional undersampling artifacts. The proposed DPS integrates a two-dimensional diffusion model, acting on image slices, coupled to sCT data consistency and volumetric regularization terms to enable 3D reconstruction robust to noise and incomplete sampling. To reduce the computational burden of DPS, stochastic contraction with PWLS initialization was used to decrease the number of diffusion steps. The validation studies involved simulations of anthropomorphic brain phantoms with synthetic bleeds and experimental data from an sCT bench. In simulations, DPS achieved ~130% reduction of directional artifacts compared to PWLS and 30% better recovery of lesion shape (DICE coefficient). Benchtop studies demonstrated enhanced visualization of brain features in a Kyoto Kagaku head phantom. The proposed DPS achieved improved visualization of intracranial hemorrhage and brain morphology compared to conventional model-based reconstruction for the highly undersampled sCT system.***

## I. INTRODUCTION

STROKE is one of the global leading causes of death and disability, with over 12 million cases per year [1]. Drastic reduction in mortality and in the prevalence and severity of sequelae has been observed with early treatment, preferably starting at the pre-hospital stage [2]. Early treatment requires reliable and fast triage between ischemic and hemorrhagic events [3], which account for ~20% of strokes [1], [3], [4]. Such triage is currently unattainable without access to neuroimaging, with multi-detector CT (MDCT) being the frontline modality for detection of Intracranial Hemorrhage (ICH) [5], [6]. Recently, Mobile Stroke Units (MSUs) have been introduced to provide rapid access to neuroimaging [5], [7]. However, MSUs rely on miniaturized MDCT or conventional cone beam CT (CBCT) with mechanical gantries requiring large vehicles for transportation and is best suited for densely populated areas.

Lighter and more compact systems have been proposed by leveraging stationary CT (sCT) configurations with multi-x-ray source arrays (MXAs) based on cold-cathode technology – e.g., Carbon Nanotubes (CNT) [8]. Previous work showed the feasibility of using an ultra-portable sCT for ICH detection [9], [10], with a configuration including a MXA and a curved area detector arranged in a cone beam geometry suitable for point-of-care deployment in conventional emergency vehicles. However, its imaging geometry presents an unconventional sampling pattern that challenges image reconstruction [9]. Specifically, the scanner acquires a sparse (31 x-ray sources) projection dataset over a limited angular arc (160º), with the peripheral views exhibiting lateral truncation. The resulting sampling is highly nonuniform across the volume.

Numerous approaches have been proposed for reconstruction in sparse and limited sampling CT, including model-based iterative reconstruction (MBIR) with sparsity-promoting, edge-preserving regularization [11], [12], dictionary-based MBIR [13], and MBIR enforcing consistency with a prior image [14], [15]. However, MBIR with edge-preserving penalties is typically insufficient for severely limited sampling. In prior-based MBIR, as well as in dictionary-based MBIR, it is often difficult to balance fidelity to the measured projections vs. enforcement of the prior in cases where the latter diverges from the object being imaged.

Recently, promising results in reduction of insufficient sampling artifacts have been shown by exploiting (statistical) prior information from large datasets, particularly using deep learning (DL) [16], [17], [18]. DL approaches often involve Convolutional Neural Networks (CNNs) for post-processing of images reconstructed with conventional algorithms. However,

This work has been submitted to the IEEE for possible publication. Copyright may be transferred without notice, after which this version may no longer be accessible. Submitted for review on May 10, 2024. This work was supported by academic/industry collaboration with Micro-X Ltd. and the Australian Stroke Alliance.

A. Lopez-Montes (alopezm2@jhmi.edu), T. McSkimming (tmcskim1@jhmi.edu), W. Zbijewski (wzbijewski@jhu.edu) and A. Sisniega (asisnie1@jhu.edu) are with the Department of Biomedical Engineering, Johns Hopkins University, Baltimore, MD 21205 USA.

T. McSkimming, A. Skeats (askeats@micro-x.com), C. Delnooz (cdelnooz@micro-x.com) and B. Gonzales (bgonzales@micro-x.com) are with Micro-X Ltd, Adelaide, SA 5042 Australia.

T. McSkimming is also with College of Science and Engineering, Flinders University, Adelaide, SA 5042 Australia.

such methods do not explicitly enforce consistency with the acquired data, making them prone to hallucinations [19], [20]. A second category integrates prior information to guide a conventional reconstruction process. CNNs have been used to create synthetic prior images [21], [22] or, alternatively, trained to learn unrolled MBIR optimization loops [23], [24]. Generally, DL- generated synthetic priors show degraded performance for out-of-domain data, while unrolled loop formulations require complex network training [23], [24].

Diffusion-based image generative models are increasingly popular for synthesis of medical imaging. Importantly, diffusion models can be relatively easily extended to enforce data consistency during the sampling process [25], [26], [27].

Assuming a continuous model, the forward diffusion process can be represented by a stochastic differential equation (SDE):

$$d\mu = f(\mu,t)dt + g(t)dw, \quad (1)$$

where $f(\mu,t)$ is a drift term that controls the evolution of the image values ($\mu$) over diffusion time ($t$), and $g(t)$ is the noise diffusion, that scales an incremental Gaussian process ($dw$). The diffusion can be reversed via a second SDE [28]:

$$d\mu = \big(f(\mu,t) - g(t)^2 \nabla_{\mu_t} \log p(\mu_t)\big)dt + g(t)dw' \quad (2)$$

Where $\mu_t$ are the image values resulting from the forward diffusion in Eq. (1) at time $t$. The term ($\nabla_{\mu_t} \log p(\mu_t)$), is denoted score, and it is often analytically intractable. Diffusion image generation models (Fig. 1.A) approximate the score term with a neural network $\nabla_{\mu_t} \log p_t(\mu_t) \approx s_\theta(\mu_t,t)$ [25], [26]. The parameters ($\theta$) of the score-matching network $s_\theta$ are obtained by training an unsupervised synthesis using a dataset assumed to be representative of the underlying noise-free image distribution $p(\mu_0)$.

To solve inverse problems using diffusion models, it would be desirable to sample, at any point $t$ in the diffusion process, from the posterior distribution given the acquired projection data $y$, $p_t(\mu_t|y)$. This can be achieved by replacing the unconditional score in Eq. (2) with the conditional score $\nabla_{\mu_t} \log p(\mu_t|y)$ [26].

The conditional score can be approximated with a deep CNN conditioned on the projection data, $\nabla_{\mu_t} \log p(\mu_t|y) = s_\theta(\mu_t,t,y)$, which must be trained in a fully supervised fashion. This formulation has shown promising performance [29], [30] but it faces challenges common to supervised training, including diminished performance in out-of-domain scenarios and inexplicit enforcement of data consistency during sampling [26], [27]. Here, we investigate an alternative based on diffusion posterior sampling (DPS) for x-ray reconstruction involving an approximation of the conditional score that applies a likelihood data-consistency correction term to the unconditional synthesis score function [27], [31].

The proposed DPS is applied to volumetric x-ray reconstruction in severely undersampled data from the portable sCT system for ICH detection. We build on our earlier preliminary studies for 2D axial sCT reconstruction [32] to develop a comprehensive framework for 3D imaging, including strategies to enforce axial inter-slice continuity in volumetric images and to reduce computational burden via stochastic contraction. The approach is evaluated in simulation studies and in experiments on an sCT benchtop. To the best of our knowledge, this is the first time a diffusion-based image reconstruction is demonstrated on experimental x-ray data.

## II. MATERIALS AND METHODS

### A. Diffusion Posterior Sampling for stationary CT reconstruction

#### 1) Diffusion Posterior Sampling

Following the Bayes rule, the conditional score function can be reformulated as follows:

$$\nabla_{\mu_t} \log p(\mu_t|y) = \nabla_{\mu_t} \log p(\mu_t) + \nabla_{\mu_t} \log p(y|\mu_t), \quad (3)$$

where $\nabla_{\mu_t} \log p(\mu_t)$ is the unconditional score, approximated using an unconditional deep CNN $s_\theta(\mu_t,t)$ as in Section I [25], [26]. The term $\nabla_{\mu_t} \log p(y|\mu_t)$ provides a mechanism to enforce consistency to the acquired data during sampling. However, no explicit dependence exists between $y$ and $\mu_t$ at an arbitrary time point $t$, except for $t=0$, which precludes the design of a closed-form solution. Under the assumption of noise free data, a surrogate for $\nabla_{\mu_t} \log p(y|\mu_t)$ can be obtained by performing an unconditional update at each time step, followed by projection of $\mu_t$ onto the measurement subspace [26], [33].

A more stable posterior sampling approach can be obtained following the derivation in [27], [31], defining $p(y|\mu_t) \approx p(y|\hat{\mu}_{0_t})$, with $\hat{\mu}_{0_t} = E[\mu_0|\mu_t]$. The posterior mean, $\hat{\mu}_{0_t}$, can be obtained by applying Tweedie's method [25], [26], [34].

To compute $p(y|\mu_t)$ for volumetric image reconstruction, we assumed a linearized forward CT model:

$$y = A \cdot \mu_t + n, \quad (4)$$

where $A$ is the forward projector operator and $n$ is the statistical noise in the acquired data.

Using the forward model in Eq. 4 and a Gaussian approximation to the noise term, the surrogate posterior score is formulated as:

$$\nabla_{\mu_t} \log p(y|\mu_t) \approx \nabla_{\mu_t} \log p(y|\hat{\mu}_{0_t}) = -\nabla_{\mu_t} \|y - A\hat{\mu}_{0_t}\|^2 \quad (5)$$

Combining Eq. 5 with the unconditional score matching approximation, the conditional update can be stated as:

$$\nabla_{\mu_t} \log p(\mu_t|y) \approx s_\theta(\mu_t,t) - \lambda \nabla_{\mu_t} \|y - A\hat{\mu}_{0_t}\|^2, \quad (6)$$

where $\lambda$ is a voxel-wise step size that weights the data consistency and the unconditional updates. The weight was set to $\lambda = 1/A^T A \mathbb{I}_\mu$, where $\mathbb{I}_\mu$ is an identity array of the same size as $\mu$, and $A^T$ is the backprojection operator.

This work used a Variance Exploding (VE) SDE for Eq. 2, in which $f(\mu,t) = 0$ and $g(t) = \sqrt{d\sigma_t^2/dt}$, yielding $\hat{\mu}_{0_t} = \mu_t + \sigma_t^2 \cdot s_\theta(\mu_t,t)$.

#### 2) Implementation of DPS for 3D reconstruction.

To limit the size of the network, $s_\theta(\mu_t,t)$, was defined as a 2-dimensional (2D) operator, yielding an unconditional model for synthesis of axial CT slices. However, the sCT system uses a cone-beam geometry and thus the conditional score in Eq. (6) requires $\hat{\mu}_{0_t}$ and $\mu_t$ to be volumetric images. Therefore, $\hat{\mu}_{0_t}$ and $\mu_t$ were obtained by stacking a number ($n_z$) of consecutive 2D slices ($\mu_{t,z}$), following $\mu_t = \{\mu_{t,z}, z \in (1, n_z)\}$ and $\hat{\mu}_{0_t} = \{\mu_{t,z} + \sigma_t^2 \cdot s_\theta(\mu_{t,z}, t), z \in (1, n_z)\}$.

However, the enforcement of 3D data consistency in a cone beam geometry following the formulation in Section II.A.1 does not guarantee continuity across consecutive slices. Previous works [34], have proposed the use of an interslice regularization function ($R_z$) acting at every sampling step. Since the data enforcement update is performed over the current estimate of the posterior mean $\hat{\mu}_{0_t}$ (which can be seen as a noise-free estimate at time $t$) and not over the evolving solution $\mu_t$, the regularization can in fact be applied not only in the inter-slice direction, but also as a more general 3D penalty. This 3D penalty promotes overall image smoothness, similar to regularization strategies commonly used in MBIR. Therefore, we use 3D regularization with an edge-preserving Huber potential commonly used in MBIR for CBCT [35]:

$$R(\hat{\mu}_{0_t}) = \sum_j \sum_{k \in N} \begin{cases} \frac{1}{2\delta}(\hat{\mu}_{0_t,j} - \hat{\mu}_{0_t,k})^2, |\hat{\mu}_{0_t,j} - \hat{\mu}_{0_t,k}| \leq \delta \\ |\hat{\mu}_{0_t,j} - \hat{\mu}_{0_t,k}| - \frac{\delta}{2}, |\hat{\mu}_{0_t,j} - \hat{\mu}_{0_t,k}| > \delta \end{cases}, \quad (7)$$

where the index $j$ runs along all voxels in the volume, and $k$ runs along the $N$ nearest neighbors of voxel $j$. In this work $N$ was set to the six nearest neighbors, two for each of the three dimensions in the volume ($v_x, v_y, v_z$). Assuming the VE SDE diffusion model in Section II.A.1, the final reverse SDE for DPS reconstruction is given by:

$$d\mu = -\frac{d\sigma_t^2}{dt}\left(s_\theta(\mu_t, t) - \lambda \nabla_{\mu_t}\left(\left\|y - A\hat{\mu}_{0_t}\right\|^2 - \beta R(\hat{\mu}_{0_t})\right)\right)dt + \sqrt{\frac{d\sigma_t^2}{dt}} dw', \quad (8)$$

where $\beta$ is a weight controlling the contribution of the penalty term to the final update. Note that $\beta$ is not necessarily a scalar and can be modified to achieve different regularization strengths in different spatial directions or to achieve spatially variant regularization with voxel-dependent $\beta$ maps. To reflect the two roles of the regularization, namely (i) enforcing inter-slice continuity in $z$, and (ii) promoting overall image smoothness; we independently varied the weight of the penalty acting in inter-slice ($\beta_z$) and in intra-slice directions ($\beta_{xy}$).

### 3) Sampling discretization and implementation

The DPS method for reconstruction of sCT is summarized in Algorithm 1 and Fig. 1.

To find a numerical solution for Eq. 8, the data consistency term (bottom row, Fig. 1B) and the noise and unconditional update terms (top row, Fig. 1B) can be computed independently at each time step, $t$. Following previous work for generative models, the unconditional terms in Eq. 8 can be discretized and solved using a Predictor-Corrector (PC) approach [25]. Following [25], we used a reverse diffusion solver for the predictor (line 6 of Algorithm 1), where $\Delta \sigma_t^2$ is the discretized version of the continuous time-dependent noise variance differential, $d\sigma_t^2/dt$. The corrector was based on annealed Langevin Markov Chain Monte Carlo (MCMC) dynamics, using the expression in line 7 of Algorithm 1, where $\varepsilon$ is a step size for the corrector update [25].

The approximation to $\nabla_{\mu_t} \log p(y|\mu_t)$ in Eq. 6 was implemented using the expression in line 10 of Algorithm 1. As illustrated in Fig. 1B, the update term was computed as the composition of one forward and one backward projection operation, multiplying the time-dependent gradient of the posterior mean $\nabla_{\mu_t}\hat{\mu}_{0_t}$. The gradient of the posterior mean was approximated as:

$$\nabla_{\mu_t}\hat{\mu}_{0_t} \approx 1 + \nabla_{\mu_t} s_\theta(\mu_t, t), \quad (9)$$

where $\nabla_{\mu_t} s_\theta(\mu_t, t)$ was computed using gradient backpropagation across the score-matching network.

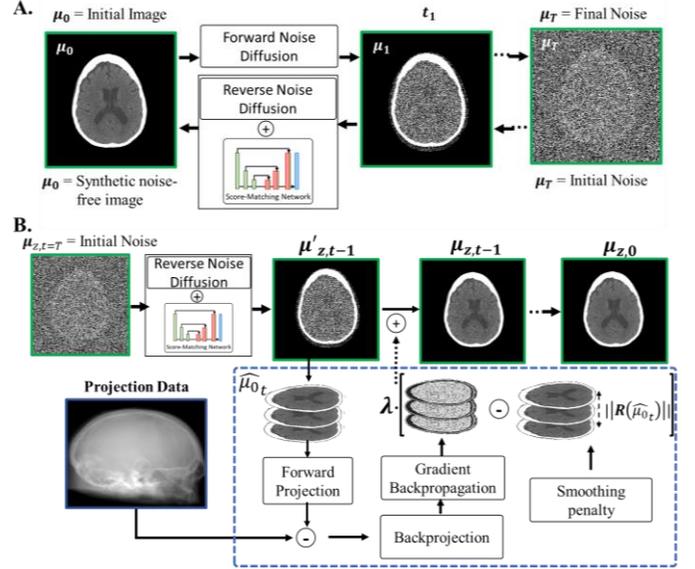

**Fig. 1.** (A) Illustration of the diffusion model for unconditional image synthesis. In forward diffusion, an initial noise-free image is corrupted by increasing levels of Gaussian noise. The process is reversed via a score-matching denoising network. (B) Schematic representation of 3D-DPS. The denoising step, guided by the score-matching network, is performed in consecutive 2D axial slices at every sampling step. Posterior sampling was achieved via data consistency and regularization (dashed box).

| **Algorithm 1. VE-SDE DPS** |
|---|
| 1. Perform MBIR $\hat{\mu}_0 \leftarrow \text{MBIR}(y)$ |
| 2. Initialize $\mu_t \leftarrow \mu_{t_{init}} = \hat{\mu}_0 + \sigma_{t_{init}} \cdot w \rightarrow w \sim N(0,1)$ |
| 3. For t = $t_{init}$: 0 do |
| 4.    For z = 1: $n_z$ do |
| 5.       $\nabla \mu_t \log p_t(\mu_{t,z}) \leftarrow s_\theta(\mu_{t,z}, t) = s_{\theta_{t,z}}$ |
| 6.       Predictor: $\mu'_{t,z} \leftarrow \mu_{t,z+1} + \Delta\sigma_t^2 s_{\theta_{t,z}} + \sqrt{\Delta\sigma_t^2} \cdot w$ |
| 7.       Corrector: $\mu'_{t,z} \leftarrow \mu'_{t,z} + \varepsilon s_{\theta_{t,z}} + \sqrt{2\varepsilon} \cdot w$ |
| 8.       Posterior mean: $\hat{\mu}_{0_{t,z}} = \mu'_{t,z} + \sigma_t^2 \cdot s_{\theta_{t,z}}$ |
| 9.    Build 3D volumes $\rightarrow \hat{\mu}_{0_t}$ and $\mu'_t$ |
| 10.   Data: $\nabla_{\mu_t}\left\|y - A\hat{\mu}_{0_t}\right\|^2 = 2A^T(y - A\hat{\mu}_{0_t})\nabla_{\mu_t}\hat{\mu}_{0_t}$ |
| 11.   Regularization: $\nabla_{\mu_t}R = \nabla_{\hat{\mu}_{0_t}}R(\hat{\mu}_{0_t}) \cdot \nabla_{\mu_t}\hat{\mu}_{0_t}$ |
| 12.   $\mu_t \leftarrow \mu'_t - \lambda\left[\nabla_{\mu_t}\left\|y - A\hat{\mu}_{0_t}\right\|^2 + \beta\nabla_{\mu_t}R\right]$ |
| 13. Return $\mu_0$ |

DPS reconstruction requires sampling of the complete diffusion path (from time $t = T$), often involving thousands of sampling steps. Following previous work on stochastic contraction for acceleration of diffusion sampling for inverse problem solving [36], the sampling process was performed along a partial diffusion path. The partial diffusion path was initialized with a single forward diffusion step acting on a conventional MBIR image $\hat{\mu}_0$, yielding $\mu_{t_{init}} = \hat{\mu}_0 + \sigma_{t_{init}} w$, with $w$ sampled form a normal distribution $\mathcal{N}(0, I)$, and $0 < t_{init} < T$. For sCT, $\hat{\mu}_0$, while contaminated by artifacts, offered a reasonable approximation to the target reconstruction, $\mu_0$.

### B. Stationary CT for detection of intracranial hemorrhage

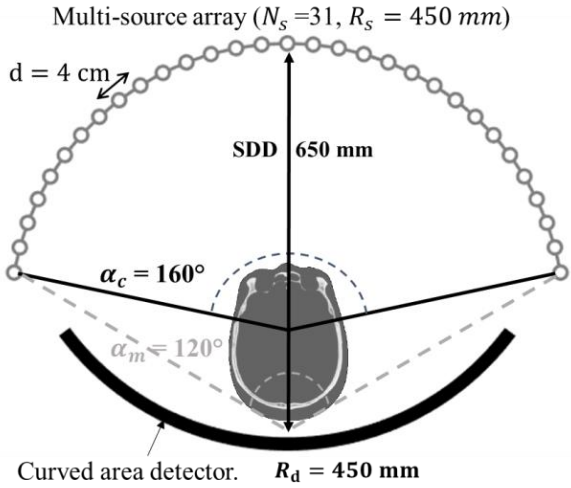

**Fig. 2.** Overview of the stationary CT configuration. An MXA, with $N_s$ = 31 CNT sources, is placed opposite to a curved area detector. Individual sources are separated by a (linear) distance of 40 mm, covering an angular span of $\alpha_c$ = 160° at the center of the FOV. The distance from the central source to the center of the detector (SDD) is 650 mm.

The concept sCT system considered in this work is illustrated in Fig. 2. The scanner is based on an MXA consisting of 31 CNT sources distributed along a circular arc with radius $R_s$ = 450 mm and centered at the center of the field-of-view (FOV). Individual sources in the MXA are evenly separated by a linear distance of $d_s$ = 40 mm. The system integrates a curved area detector with a curvature radius of $R_d$ = 450 mm. The source-to-detector distance (SDD), defined between the center of the detector and the central MXA element, is 650 mm. The detector is 1650 x 600 pixels with an isotropic pixel size of 0.5 mm.

Compared to conventional CBCT, the sampling in sCT is challenged by: (i) sparsity, since only 31 views are acquired; (ii) limited angular coverage – at the center of the FOV, the MXA spans $\alpha_c$ = 160°, which is less than 180° + fan angle; and (iii) nonuniformity – e.g., the angular span at posterior regions of the FOV (the closest to the detector) drops to $\alpha_p$ = 120°.

Two sCT protocols were considered: (i) a fully stationary scan, which suffers from sparse, limited, and nonuniform angular sampling as described above; and (ii) a semi-stationary scan, which combines two stationary acquisitions (31 projection views/each) separated by a 45° pivoting motion executed around the superior-inferior axis of the FOV. The latter provides an upper sCT performance bound since it yields a more complete and uniform angular sampling ($\alpha_c$ = 205°) than the former, but remains sparse, which is anticipated for a realistic, economical configuration of stationary emitters.

### C. Assessment of DPS in sCT brain imaging

#### 1) The unconditional brain CT synthesis model

Following [25], we trained a Noise Conditional Score Network (NCSN++) minimizing the following loss function:

$$\theta^* = argmin_\theta \sum_t \sigma_t^2 \mathbb{E}[\mu_0] \mathbb{E}[\mu_t|\mu_0] [||s_\theta(\mu_t, t) - \nabla_{\mu_t} \log p_{0_t}(\mu_t|\mu_0)||^2], \quad (10)$$

where $p_{0_t}(\mu_t|\mu_0) = \mathcal{N}(\mu_t|\mu_0, \sigma_t^2)$.

The diffusion model was trained using brain CT images from the publicly available Head CT CQ500 dataset [37]. The training set consisted of a random selection of 250 MDCT volumes, combining cases with and without ICH. The network was trained on 2D axial slices, with 100 slices extracted randomly from each volume (total of 25000). Training was performed in an unsupervised fashion for 500 epochs, with the Adam optimizer and 2000 steps for reverse sampling. No data consistency was included in the training.

#### 2) Simulation studies

Validation in a simulation study involved a set of digital head phantoms obtained from 11 MDCT volumes acquired in-house in a previous IRB-approved study. Projections of the phantoms were obtained using a high-fidelity CT simulator [21] assuming a polychromatic x-ray spectrum (100 kV + 2 mm Al filtration) and realistic acquisition protocols with increasing exposure: 0.125, 0.25, 0.5 and 1 mAs per frame. The expected exposure for the physical sCT scanner is 0.5 mAs per projection, corresponding to 1 mGy dose to soft-tissue at the isocenter, estimated from a Monte Carlo simulation [38], for the stationary case (2 mGy for the semi-stationary). The x-ray spectrum, tube output per mAs, and material attenuations were obtained from Spektr [39]. Quantum and electronic noise were introduced using a previously validated model including noise correlations [21] parameterized from experimental measurements of flat-panel detectors for head CBCT [40].

To evaluate DPS for ICH visualization using sCT, randomly-shaped simulated bleeds of 40 HU contrast and volume ranging from 0.5 ml to 17 ml were inserted into the brain tissue of the head phantoms. Lesion contrast was representative of blood-to-brain parenchyma [41] and their size spanned the range from subclinical bleeds to extended ICH [42], [43]. Each bleed was generated from a spherical base shape by application of a random deformation field (i.e., the directions and amplitudes of the field were randomized at each spatial location.)

Nine lesions were inserted into each test volume. The brain was conceptually divided into three broad regions of interest along the superior-inferior axis: (i) superior (>15 mm superiorly from the center of mass of the brain), characterized by fine details such as the sulci; (ii) central (<15 mm superior and inferior from the center of mass), containing the ventricles; and (iii) inferior (>15 mm inferior from center of brain), where intricate bony anatomy tends to exacerbate sampling artifacts. Three synthetic bleeds were placed in each of the three regions, at random locations within the region; one mimicked a subclinical ICH < 2.0 ml, one represented a moderate ICH > 2.0 ml and < 7.0 ml, and one presented an extensive ICH > 7.0 ml.

The complete dataset included 99 bleeds: 11 base anatomies, three ICH sizes, and three anatomical regions.

Image reconstructions were obtained using DPS and a conventional Penalized Weighted Least Squares (PWLS) MBIR, each performed on a 256 x 256 x 175 voxels grid (1 mm isotropic). In DPS, we applied stochastic contraction and investigated the effects of changing the number of sampling steps, ranging 100-2000, where 2000 was the diffusion sequence length used to train the unconditional synthesis. Further, we compared inter-slice only regularization ($\beta_z = 2$, $\beta_{xy} = 0$) to 3D regularization ($\beta_{xy} = 1$ and $\beta_z = 2$), both with $\delta = 5 \cdot 10^{-4}\ mm^{-1}$. The reference PWLS also employed the 3D edge-preserving Huber penalty, which is commonly used in sparse sampling scenarios [44]. The settings were $\beta = 2$ isotropic, $\delta = 5 \cdot 10^{-4}\ mm^{-1}$. Two hundred iterations of PWLS with Nesterov momentum acceleration were performed [45], [46]. Following previous work on this sCT configuration [9], a tapering window (erf function with $\sigma_{erf} = 10$ mm) was used to reduce the effect of lateral truncation in peripheral sources.

Overall image quality was evaluated in terms of the structural similarity index (SSIM) between a reconstructed volume ($R$) and the ground truth ($GT$) digital phantom (MDCT with the inserted lesions).

Recognizing the directional nature of sampling artifacts in sCT, their conspicuity was further assessed in terms of the alignment of directional image gradients between $R$ and $GT$. This was quantified using Tamura directionality textures [47]:

$$\Delta D = \frac{100}{n_b} \cdot \sum_{n_b} \frac{|h(\alpha_{|\nabla\mu|>\epsilon})|_R - h(\alpha_{|\nabla\mu|>\epsilon})|_{GT}|}{h(\alpha_{|\nabla\mu|>\epsilon})|_{GT}}, \quad (11)$$

where $h(\alpha_{|\nabla\mu|>\epsilon})$ is a smoothed histogram of gradient orientations in an image slice (64 equally spaced angular bins ranging 0-180° followed by Gaussian smoothing with $\sigma = 3$ bins; 0° corresponds to the axis connecting the FOV center to the center of the detector). Only gradients with magnitude larger than $\epsilon = 6 \cdot 10^{-5}$ mm$^{-1}$ were considered. For each $R$ and $GT$ pair, $\Delta D$ was computed per slice and aggregated by brain region (inferior, central, and superior) for analysis. We also report averages of the slice measurements across the entire brain and across all test volumes, denoted $\overline{\Delta D}$, as a function of scan and reconstruction settings.

The fidelity of the reconstructed bleeds was assessed using DICE similarity coefficient between a segmentation of the bleed in a reconstructed volume (binary mask $M_R$) and its ground truth shape ($M_{GT}$). The segmentation was achieved by thresholding with lower and upper thresholds of 20 HU and 120 HU, respectively, performed within a square ROI with side equal to twice the nominal diameter of the lesion, centered at the ground truth location of the lesion.

*3) Experimental studies*

Experimental evaluation was performed using a benchtop system emulating the proposed sCT geometry (Fig. 3). The bench integrated a curved amorphous Silicon (aSi) area detector (Fujifilm, Japan) with $R_d = 486$ mm. The detector was built from two modules of 425 x 425 mm², the pixel size was 0.15 mm, the scintillator was a 0.3 mm thick GdOS. A single CNT-based x-ray source (Mini-CNT, Micro-X, Australia) was mounted on a robotic positioning system (BiSlide, Velmex Inc., NY) and stepped on an arc ($R_s = 450$ mm, $SDD = 650$ mm) to simulate the 31-view stationary MXA protocol of Sec. B.1. The object was placed on a mechanized rotary table (B48 Series, Velmex Inc., NY); the semi-stationary acquisition was achieved by rotating the object by 45° between two source trajectory sweeps.

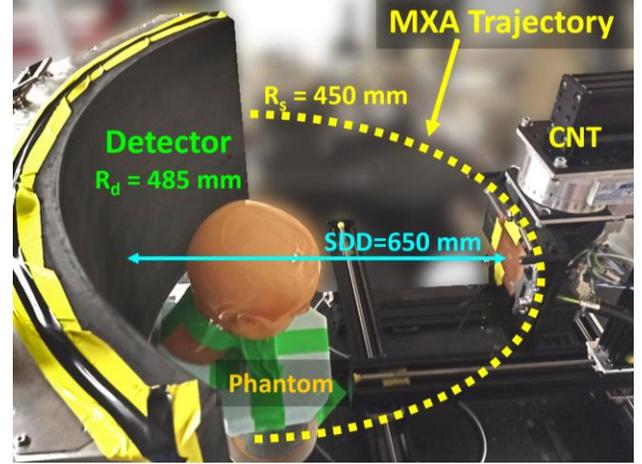

**Fig. 3.** X-ray test bench emulating the proposed stationary CT (sCT). The source trajectory for simulation of the 31-element MXA is marked in yellow.

An anthropomorphic brain phantom (Kyoto Kagaku Ltd., Kyoto, Japan) was imaged using the stationary and semi-stationary protocols. The x-ray technique was 102 kV (+2 mm Cu filtration) and 0.5 mAs/projection. Prior to reconstruction, raw projection data were flat-field corrected, followed by projection-domain scatter compensation [10] and log conversion. PWLS and DPS reconstructions were performed on a 256 x 256 x 200 voxel grid (1 mm isotropic). The hyperparameter settings were the same as in simulations. An MDCT volume of the phantom provided the ground truth for evaluation.

### III. RESULTS

*A. ICH imaging performance of PWLS and DPS in simulated sCT.*

This section summarizes the performance of DPS and PWLS using simulation data for a nominal sCT protocol with 0.5 mAs/projection. The DPS settings are fixed to inter-slice only penalty ($\beta_z = 2$, $\beta_{xy} = 0$), and stochastic contraction with 700 time-steps, as per optimization studies in section III.B. Fig. 4 shows example DPS and PWLS reconstructions for the stationary (31 views, $\alpha_c = 160°$) and semi-stationary (62 views, $\alpha_c = 205°$) sCT configurations and the corresponding SSIM maps. Compared to PWLS, DPS yielded a reduction of sampling artifacts and improved visualization of the simulated bleeds and brain parenchyma in both protocols. The shape of the lesions was recovered more accurately with DPS. Particularly notable, especially in the stationary protocol, is a substantial improvement in SSIM of the brain structures in the inferior region of the skull, where the sampling artifacts are most pronounced due to the complex bony anatomy.

For sCT, DPS yielded a mean SSIM, averaged across the inferior region of the brain, of 0.93, compared to 0.84 for PWLS. Illustrating the better image quality at superior and central regions of the brain, DPS obtained a mean SSIM of 0.96 for those locations. This larger similarity was however a more modest improvement compared to the PWLS baseline mean SSIM of 0.91. Similarly, when using the better-posed semi-stationary protocol, mean SSIM values were higher with PWLS (0.94 for inferior, and 0.96 for superior and central locations), and the improvement obtained with DPS was comparatively lower, resulting in a mean SSIM of 0.96 for inferior brain regions, and 0.97 for superior and central locations.

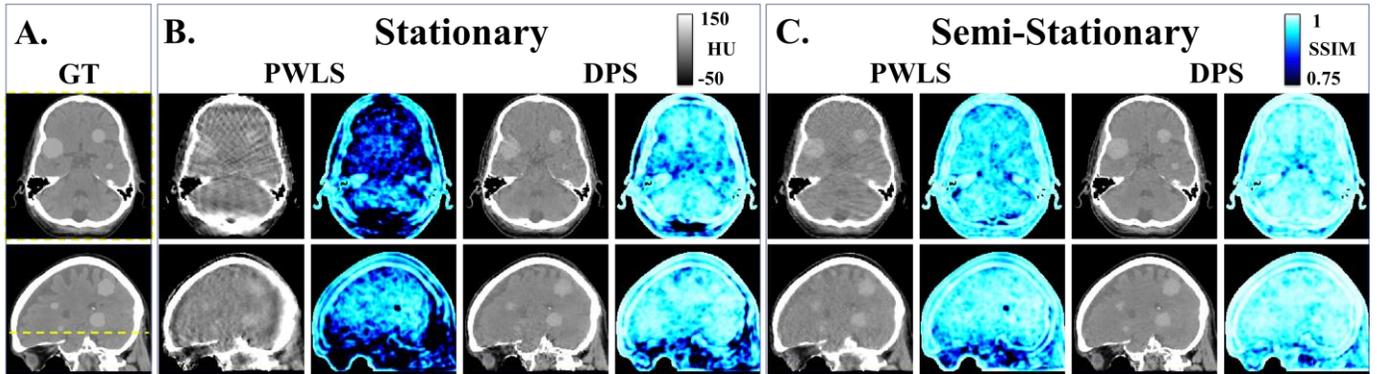

**Fig. 4.** Example reconstructions of simulated head (A) data obtained at 0.5 mAs/projection using the stationary sCT protocol (B) and the semi-stationary protocol (C). SSIM against the phantom is shown for each reconstructed image. Regularization settings for DPS were $\beta_z = 2$, $\beta_{xy} = 0$.

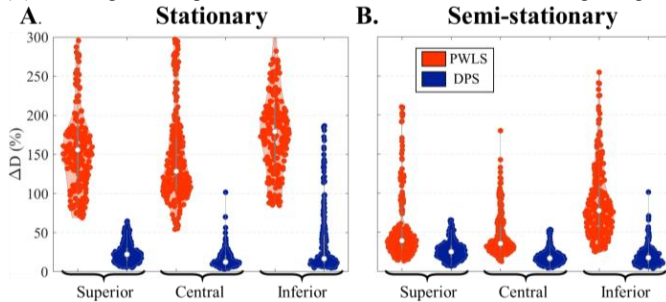

**Fig. 5.** Directional content agreement ($\Delta D$) between the ground truth and PWLS and DPS reconstructions. Each data point represents a slice in one test case reconstruction (A – stationary sCT, B – semi-stationary).

Fig. 5 demonstrates that the artifact reduction observed in DPS is reflected in better agreement of directional image content with the ground truth. In the challenging stationary protocol (Fig. 5A), $\overline{\Delta D}$ is 26% for DPS compared to 175% for PWLS. In the semi-stationary scan with better geometric sampling (Fig. 5B), the performance of PWLS is improved but still affected by residual artifacts ($\overline{\Delta D}$ = 62%), whereas DPS appears to provide a similar degree of artifact mitigation in both protocols (semi-stationary $\overline{\Delta D}$ = 21%). Consistently with Fig. 4, the inferior brain region shows the most pronounced $\Delta D$ decrease from PWLS to DPS.

The DICE for each simulated bleed is shown in Fig. 6; the data is stratified by brain region and lesion volume. In the stationary protocol, accurate recovery of the ICHs is challenging with PWLS, as evidenced by median DICE < 0.7 for < 2 ml bleeds and < 0.8 for > 2 ml bleeds, placed in the superior and central brain regions. The performance is worse at inferior locations, particularly for < 2 ml bleeds (median DICE ~0.6), consistent with more severe artifacts in Fig. 4. In comparison, DPS achieves median DICE > 0.8 for > 2 ml ICH across the entire brain, DICE > 0.75 for < 2ml lesions in the central and superior regions, and DICE of ~0.7 for < 2 ml bleeds in the inferior region. As anticipated, ICH recovery in the semi-stationary protocol is more robust using both PWLS and DPS (DICE > 0.8 across anatomical locations and sizes) than with the fully stationary protocol due to less severe undersampling.

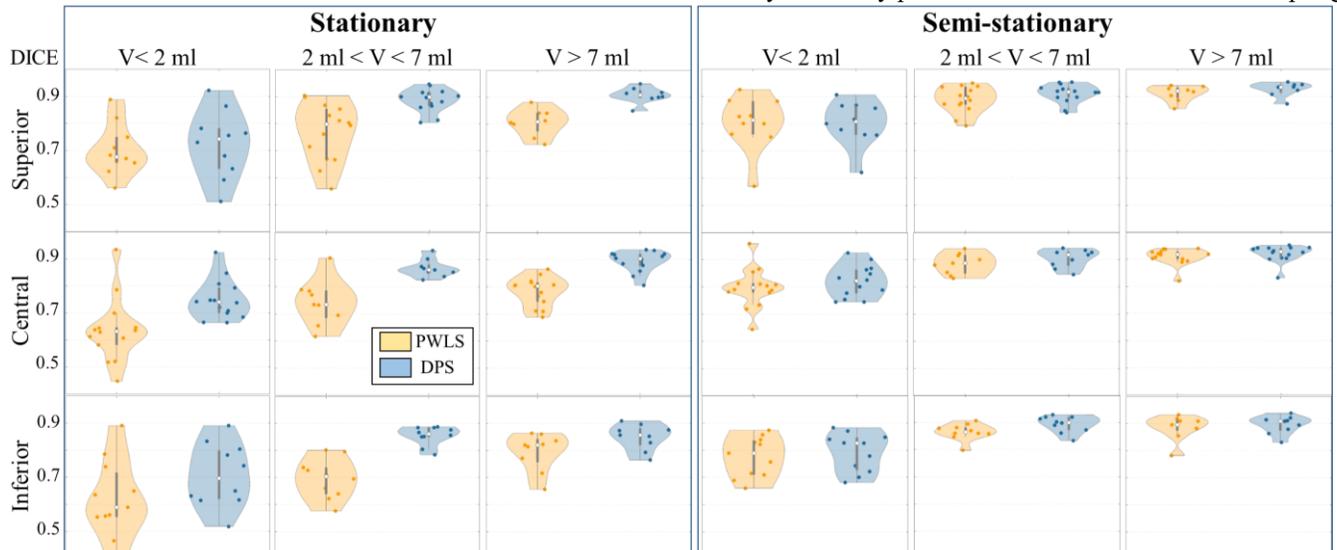

**Fig. 6.** DICE scores of segmented ICHs. (A – stationary sCT, B - semi-stationary). Each data point is one bleed in one of the 11 test heads. Lesions are stratified by volume (V) and location in the brain parenchyma. PWLS results shown in yellow (left) and DPS is in blue (right).

## B. Stochastic sampling and regularization considerations in DPS.

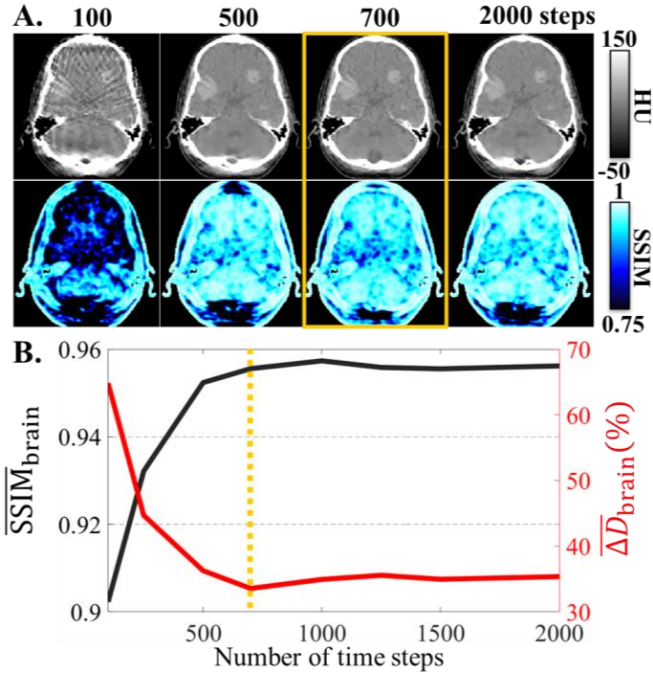

**Fig. 7.** Performance of DPS with stochastic contraction, initialized using PWLS, as a function of the number of reverse diffusion time steps after initialization. (A) Reconstructed images for a representative test case. (B) SSIM and directional content agreement averaged over the brain region.

This section uses the simulated stationary sCT data (31 views, $\alpha_c = 160°$) to examine: (i) the impact of stochastic contraction with PWLS initialization on DPS performance; (ii) the stochastic variability of the posterior sampling outcome; and, (iii) the impact of noise and regularization on DPS.

Fig. 7 illustrates the effect of using PWLS reconstruction to initialize the reverse diffusion for different numbers of diffusion steps (i.e., different degrees of stochastic contraction). The initial PWLS reconstruction was obtained using the setup and parameterization described in Section II.C.2. Note that increasing the number of steps implies that more noise is injected at the start of the diffusion (larger $\sigma_{t_{init}}$ in Algorithm 1); for the maximum $t_{init} = T = 2000$ steps, the PWLS image after noise injection is indistinguishable from pure Gaussian noise. Reconstructed images (A) and $\Delta D$ and SSIM averaged across the brain (B) for a representative test case show that at least ~500 steps are needed to build a diffusion path capable of compensating sampling artifacts in the PWLS initialization. In light of the trends in Fig. 7, minimal improvements in image quality and SSIM were observed for > 700 sampling steps. Thus, all experimental studies presented throughout this work were obtained with stochastic contraction and 700 sampling steps unless otherwise noted.

Even for the same projection data, DPS with different reverse diffusion paths (random seeds) will produce different samples from the posterior distribution. Fig. 8 evaluates the variability of the reconstructed images for an example dataset using 10 realizations of DPS. Voxel-wise mean attenuation ($\overline{DPS}$) and coefficient of variation (CV) were computed across the resulting 10 reconstructions. The observed variability was limited to subtle anatomical features, primarily in the bone (<3% CV for the entire head but < 0.5% CV in brain only). The lesions are consistently recovered in all realizations, as evidenced by their sharp delineation in the mean image and low CV. The variability is generally higher in the inferior region of the head, where uncertainty from the conjunction of limited sampling and high-frequency, high-contrast features led to a higher uncertainty in the reconstructed structures.

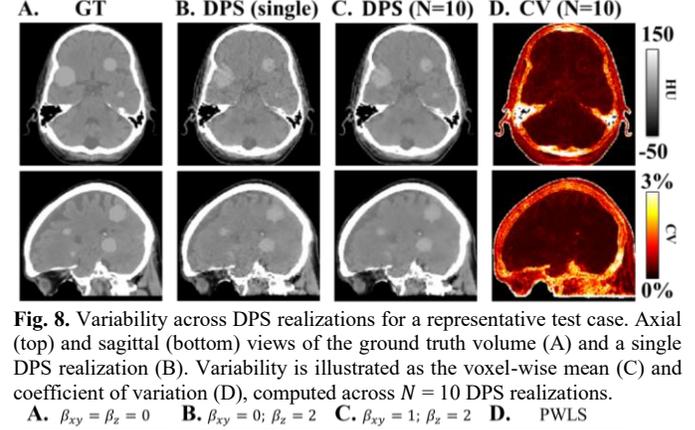

**Fig. 8.** Variability across DPS realizations for a representative test case. Axial (top) and sagittal (bottom) views of the ground truth volume (A) and a single DPS realization (B). Variability is illustrated as the voxel-wise mean (C) and coefficient of variation (D), computed across N = 10 DPS realizations.

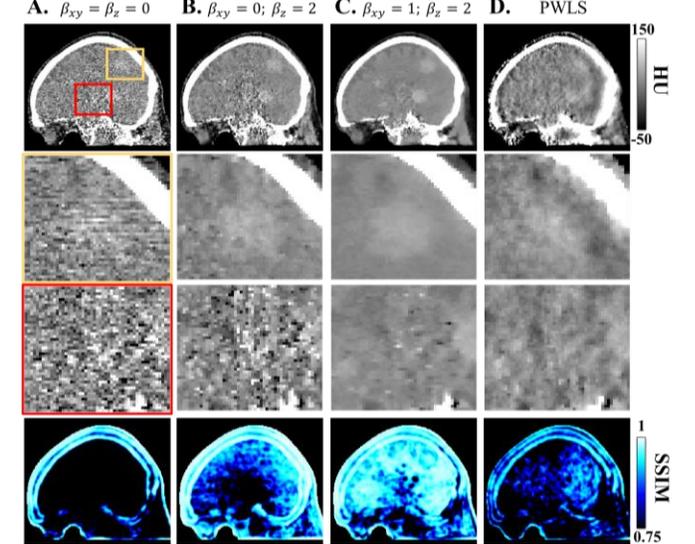

**Fig. 9.** Example images for the lowest investigated exposure level (0.125 mAs/projection) and various regularizations. Top row: example coronal slice. 2nd row: image region affected by inter-slice artifacts (indicated by an orange box in the full coronal view). 3rd row: skull-base region affected by noise (red box). Bottom row: SSIM maps for the full coronal view. (A) DPS with no regularization. (B) DPS with inter-slice only penalty. (C) DPS with inter-slice + in-slice regularization. (D) PWLS at the same exposure level.

Fig. 9 and 10 examine the performance of DPS for a range of exposure levels and different regularization strategies. Fig.9 shows DPS sagittal slices for a representative case at the lowest studied exposure (0.125 mAs/projection). Without regularization (Fig. 9A), artifacts (horizontal banding) due to inconsistencies between 2D axial slices are apparent in addition to the high overall level of image noise. This is reflected in the poor SSIM (bottom row). Inter-slice-only regularization (Fig. 9B) effectively suppressed these artifacts. However, it does not achieve sufficient noise reduction at this very low exposure, especially in the highly attenuating region of the skull base, particularly in its center. The addition of in-slice regularization (Fig. 9C) yielded improved noise control and better visualization of the lesions. Fig. 9D shows the PWLS reconstruction for the same exposure for comparison.

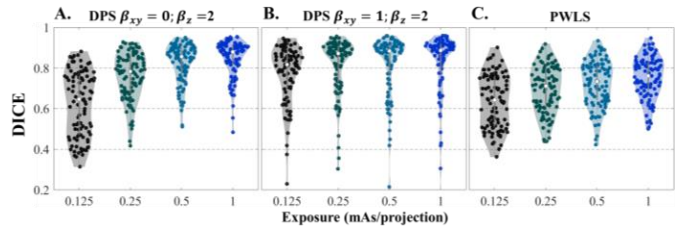

**Fig. 10.** DICE score as a function of x-ray exposure (0.125-1 mAs/projection) and regularization strategy (A - Inter-slice penalty), (B - 3D penalty) compared to PWLS results (C).

Fig. 10 summarizes DICE scores of all test lesions as a function of exposure level and DPS regularization. Inter-slice only regularization (Fig. 10A) is more sensitive to noise at low exposures (<=0.25 mAs) than 3D regularization (Fig. 10B); e.g. the median DICE score at 0.125 mAs is < 0.65 without in-slice regularization compared to > 0.8 with in-slice regularization. The increased number of outliers (DICE <0.5) for the 3D penalty compared to the inter-slice penalty is due to the increased blur, which affects the recovery of small bleeds (<2 ml). Fig. 10.C shows the DICE score for PWLS, presenting a similar trend of reduced performance with increasing noise as in DPS, and a consistently lower DICE value compared to DPS.

### C. Experimental feasibility study.

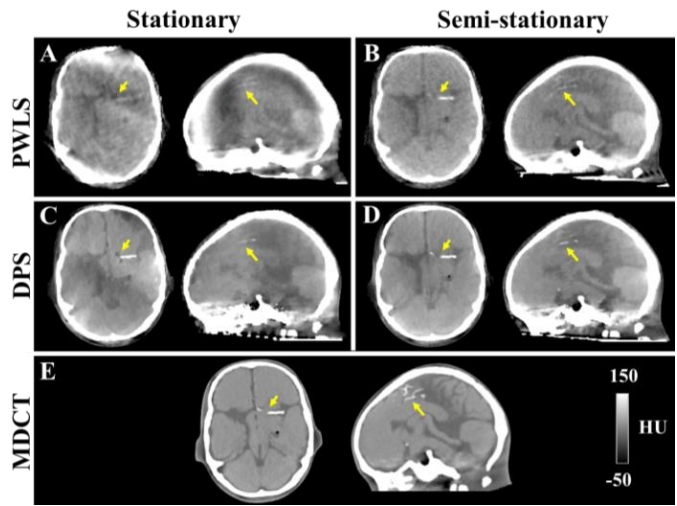

**Fig. 11.** Axial and sagittal views of volumetric reconstructions from a bench study of an anthropomorphic brain phantom. PWLS reconstructions for the stationary (A) and semi-stationary (B) protocols show artifacts and non-uniformity due to limited and sparse sampling, hampering the visibility of contrast-enhanced vessels (see arrows). DPS reconstructions for the stationary (C) and semi-stationary (D) protocols show noticeable reduction of sampling artifacts, improved recover of the vessels, and better agreement with the MDCT reference (E).

Figure 11 shows DPS reconstructions with inter-slice regularization ($\beta_z = 2$ and $\beta_{xy} = 0$) in experimental phantom data, for the stationary sCT protocol (31 views, $\alpha_c = 160°$) and for the upper bound semi-stationary protocol (62 views, $\alpha_c = 205°$). For sCT, PWLS yielded reconstructions with noticeable sampling artifacts and contrast loss, more conspicuous in frontal and posterior regions. Artifacts and non-uniformity hampered the visualization of vascular and soft-tissue structures throughout the brain (Fig. 11A). Compared to PWLS, DPS resulted in a noticeable reduction of sampling artifacts, yielding better delineation of brain midline, ventricle boundaries, and vascular structures, in both axial and sagittal views. The remaining image non-uniformity in DPS is partly attributable to residual uncorrected scatter and beam hardening that impart low-frequency attenuation inconsistencies not considered in training of the unconditional generative model.

In agreement with simulation studies, the semi-stationary protocol yielded higher baseline quality with PWLS. Effects of sparse sampling are evident as mid-frequency artifacts throughout the brain parenchyma, blurring of soft-tissue boundaries, and blurring and distortion of small vascular structures (Fig. 11B). In line with previous observations, DPS resulted in an overall higher perceived image quality, a better delineation of soft-tissue boundaries, and more conspicuous vascular features.

## IV. DISCUSSION AND CONCLUSIONS

The presented results demonstrate that DPS enables ICH detection using the proposed sCT configuration by achieving substantial reduction of sampling artifacts compared to conventional MBIR with edge-preserving regularization. Simulation studies of the stationary protocol showed accurate DPS reconstruction of synthetic bleeds with contrast pertinent to fresh blood to brain parenchyma (40 HU), and volume ranging from subclinical to extended ICH.

For large and medium-sized bleeds (>2 mL), DPS reconstructions yielded good recovery of their attenuation, shape, and spatial extent in the superior and central brain regions, as evidenced by median DICE > 0.85, compared to ~0.7 for PWLS. The increased visibility of ICH with DPS is largely attributable to a reduction of directional sampling artifacts (~150% decrease in $\Delta D$ with DPS). Consistently with this observation, the performance of DPS was slightly degraded in inferior regions of the brain, where undersampling artifacts are more conspicuous due to intricate skull base anatomy.

Compared to medium and large ICH, the visibility of small bleeds (<2 ml) was more severely challenged by undersampling in the stationary protocol. The lower visibility was reflected in a lower DICE for PWLS and a larger variability across cases, particularly for superior and inferior brain locations where artifacts from angular and cone-beam undersampling are more conspicuous. Despite the overall lower visibility of small ICH, DPS remained superior to the reference PWLS, particularly in inferior brain regions where median DICE for DPS was 0.70, compared to 0.59 for PWLS.

The difference in performance between DPS and PWLS was less marked in the semi-stationary protocol. However, even with the improved sampling provided by this configuration, DPS remained advantageous to PWLS. Directional artifacts ($\Delta D$) were reduced by ~40%, with the most conspicuous improvement in the inferior region of the brain where the baseline $\Delta D$ was generally elevated due to high-contrast skull-base features that exacerbate sampling effects.

As common to diffusion-based generative models, the stochastic sampling nature of DPS introduces variability in the

9reconstructed volume (for the same CT projection data) as a function of the random diffusion path during reverse sampling. However, the results in Section III.B show that the variation across diffusion realizations is <3% for the entire volume (0.5% for brain tissue inside the skull) and mostly restricted to low-contrast brain features such as small sulci and ventricle boundaries. It was elevated in posterior brain regions, illustrating how the fraction of the posterior distribution that is consistent with measured projections expands when the geometric sampling is severely limited, as is the case in these locations. Nevertheless, this observed variability was found to have minimal impact on ICH visibility. For applications of DPS in other neuroimaging scenarios, the stochastic variation must be separately evaluated for the pertinent imaging tasks.

The proposed volumetric smoothness penalty illustrates the possibility of merging conventional regularization with learned generative models for posterior sampling. A noticeable reduction of noise-induced image quality degradation in low-dose scenarios was achieved. Further, volumetric continuity was effectively enforced, enabling the use of a two-dimensional score-matching network for fully volumetric reconstruction. The impact of regularization on spatial resolution was assessed visually and indirectly using ICH DICE metrics; further quantitative evaluation as a function of imaging task [44], [48] is a subject of ongoing work.

To the best of our knowledge, the experimental benchtop studies presented in this paper are the first demonstration of diffusion-based volumetric tomographic reconstruction in physical CT data. Despite the unavoidable deviations in signal and noise properties between the simulations and the benchtop configuration (e.g., x-ray beam spectrum, detector response, noise, or residual biases in projection data), DPS outperformed conventional MBIR in terms of visual image quality and visibility of brain anatomy and contrast-enhanced vessels. Consistently with the robustness to projection noise illustrated in section III.B, the results suggest a minimal impact of any mismatches between the simulated and experimental noise on the perceived DPS image quality. However, a low-frequency projection bias due to residual uncorrected scatter resulted in shading artifacts in DPS, particularly conspicuous for the limited sampling scenario of the stationary protocol. The impact of such bias on DPS image quality is a subject of ongoing research. Further, the experimental testbench featured a two-module curved panel detector with unsampled (dead) central columns. Compared to the continuous detector in the simulations, this discontinuity exacerbated the heterogeneity of the volumetric sampling, challenging both DPS and PWLS.

A limitation inherent to generative diffusion models is the high computational burden of step-wise sampling of the reverse diffusion path. The results presented here were obtained with an unoptimized implementation on a single GPU (Nvidia RTX A6000). For this setup, the reconstruction time was ~12 hours for a 256x256x175 voxels volume using 2000 sampling steps. A naïve acceleration could be achieved by simple reduction of the number of diffusion steps. However, this has been shown to substantially degrade the generated images [49]. Instead, we applied stochastic contraction, achieving comparable image quality with only 700 steps and a reconstruction time of 4 h (65% less than without contraction). Preliminary results show that configurations with denser volumetric sampling might allow for further acceleration – for example, for the semi-stationary protocol discussed here, 250 sampling steps appear sufficient using stochastic contraction.

Further reduction in computational burden could be achieved with alternative approaches to sampling the reverse diffusion path that require a lower number of steps compared to the scheme used here. Examples include adaptive schedules for the sampling step [49], [50], and geometric decomposition of the diffusion path [51]. Alternatively, substantial acceleration has been demonstrated with latent-space methods, in which the diffusion generative model is applied to a latent representation of reduced dimensionality obtained by a trained CNN [52]. While the relation between the latent representation and the image space is not strictly linear, preliminary work has shown the feasibility of this approach in inverse problems [53].

In summary, DPS is a promising approach that outperforms conventional MBIR for tomographic reconstruction in stationary CT with limited, sparse, and inhomogeneous volumetric sampling. The application of DPS is a potential paradigm change that may enable point-of-care imaging with highly portable devices based on simple hardware.